\pdfoutput=1
\documentclass{Interspeech}

\usepackage[caption=false,font=normalsize,labelfont=sf,textfont=sf]{subfig}
\usepackage{textcomp}
\usepackage{stfloats}

\usepackage{xparse}
\usepackage{amsmath}
\usepackage{commath}
\usepackage{amssymb}
\usepackage{mathtools}
\usepackage{bbm}
\usepackage{physics}
\usepackage{siunitx}
\usepackage{trfsigns}
\usepackage{pifont}
\usepackage{xspace}
\usepackage{etoolbox}
\usepackage{multirow}
\usepackage{csquotes}
\usepackage{graphicx}
\usepackage[nosort]{cite}   % Groups citations like [1-4] instead of [1, 2, 3, 4]

\usepackage{tablefootnote}

\usepackage{hyperref}
\usepackage[capitalize]{cleveref}
\usepackage [table]{xcolor}
\usepackage{balance} % For the balanced reference page
\usepackage{algpseudocode}
\usepackage{algorithm}
\usepackage{booktabs}

\usepackage{url}

\newcolumntype{H}{>{\setbox0=\hbox\bgroup}c<{\egroup}@{}}   % Hidden column https://tex.stackexchange.com/a/16607/148912
\NewDocumentCommand{\tab}{ O{c} O{c} m }{\begin{tabular}[#2]{@{}#1@{}}#3\end{tabular}}   % Simple tabluar that does not add vertical spacing
\newcommand*{\thl}{\fontseries{b}\selectfont}
\robustify\thl
\robustify\fontseries
\robustify\underline

\definecolor{tikz_orange}{HTML}{ee854a}
\definecolor{tikz_green}{HTML}{6acc64}
\definecolor{tikz_red}{HTML}{d65f5f}
\definecolor{tikz_violet}{HTML}{956cb4}
\definecolor{tikz_gray}{HTML}{797979}
\definecolor{muted8}{HTML}{d5bb67}
\definecolor{tikz_teal}{HTML}{82c6e2}
\usepackage{pgfplots} % Import of plots from tikzplotlib
\usepgfplotslibrary{groupplots}
\usepackage{tikzsymbols}
\usepackage{pgfplotstable}

\newlength\fheight % Plots from Matlab share the same size
\newlength\fwidth % Plots from Matlab share the same size
\newlength{\figureheight} % Plots from Matlab share the same size
\newlength{\figurewidth} % Plots from Matlab share the same size

\usepackage{tikz}
\pgfplotsset{compat=1.9}

\usetikzlibrary{external}
\usetikzlibrary{arrows}
\usetikzlibrary{patterns}
\usetikzlibrary{backgrounds}
\usetikzlibrary{fit}
\usetikzlibrary{positioning}
\usetikzlibrary{shapes.geometric}
\usetikzlibrary{calc}
\usetikzlibrary{shapes.multipart}
\usetikzlibrary{shapes.misc}
\usetikzlibrary{shapes.symbols}
\usetikzlibrary{patterns.meta}
\usetikzlibrary{fadings}

\pgfdeclarelayer{bg}    % declare background layer
\pgfsetlayers{bg,main}  % set the order of the layers (main is the standard layer)

\tikzset{>=stealth}

\tikzstyle{block}=[
    draw,
    text depth=0pt,
    thick, 
    rectangle,
    text centered,
    minimum height=3.4ex,
    rounded corners=0.3em,
    fill=black!6,
    inner sep=0.6em,
    ] % rectangle

\tikzstyle{dnn}=[]
\tikzstyle{enhBlock}=[]%fill=red]
\tikzstyle{estBlock}=[dashed]%, fill=green]

\tikzstyle{branch}=[{circle,inner sep=0pt,minimum size=0.3em,fill=black}]
\tikzstyle{box}=[rectangle, rounded corners, draw=black, line width=1pt, text width=2cm]

\tikzstyle{arrow}=[{}-{>}, thick]
\tikzstyle{line}=[thick]
\tikzstyle{reverse arrow}=[{<}-{}, thick]

\tikzset{% http://tex.stackexchange.com/a/257632
	do path picture/.style={%
		path picture={%
			\pgfpointdiff{\pgfpointanchor{path picture bounding box}{south west}}%
			{\pgfpointanchor{path picture bounding box}{north east}}%
			\pgfgetlastxy\x\y%
			\tikzset{x=\x/2,y=\y/2}%
			#1
		}
	},
	sin wave/.style={do path picture={    
			\draw [line cap=round] (-3/4,0)
			sin (-3/8,1/2) cos (0,0) sin (3/8,-1/2) cos (3/4,0);
	}},
	cross/.style={draw, circle, do path picture={    
			\draw [line cap=round] (-2/5,-2/5) -- (2/5,2/5) (-2/5,2/5) -- (2/5,-2/5);
	}},
	plus/.style={draw, circle, do path picture={    
			\draw [line cap=round] (-3/5,0) -- (3/5,0) (0,-3/5) -- (0,3/5);
	}},
	mic/.style={inner sep=0pt, do path picture={
			\draw (0,0) circle (0.9);
			\draw [line cap=round] (-0.9, -0.9) -- (-0.9, 0.9);
	}},
	mux/.style={trapezium, draw}
}

\usepackage[acronym,shortcuts]{glossaries}

\glsdisablehyper

\newacronym{STFT}{STFT}{Short-Time Fourier Transform}
\newacronym{DER}{DER}{Diarization Error Rate}
\newacronym{WER}{WER}{Word Error Rate}
\newacronym{SRO}{SRO}{Sampling Rate Offset}
\newacronym{STO}{STO}{Sampling Time Offset}
\newacronym{EEND}{EEND}{End-to-End Neural Diarization}
\newacronym{TDOA}{TDOA}{Time Difference Of Arrival}
\newacronym{GCC-PhaT}{GCC-PhaT}{Generalized Cross Correlation with Phase-Transform}
\newacronym{cpWER}{cpWER}{concatenated minimum-permutation Word Error Rate}
\newacronym{SCM}{SCM}{Spatial Covariance Matrix}
\newacronym{GSS}{GSS}{Guided Source Separation}
\newacronym{ASR}{ASR}{Automatic Speech Recognition}
\newacronym{cACGMM}{cACGMM}{complex Angular Central Gaussian Mixture Model}
\usepackage[symbol]{footmisc}
\usepackage{hyperref}

\interspeechcameraready

\title{Spatio-spectral diarization of meetings by combining TDOA-based segmentation and speaker embedding-based clustering}
\author[equalcontribution]{Tobias}{Cord-Landwehr}
\author[equalcontribution]{Tobias}{Gburrek}
\author[]{Marc}{Deegen}
\author[]{Reinhold}{Haeb-Umbach}

\affiliation[nocounter]{Paderborn University}{Communications Engineering Department}{Germany}

\email{\{cord,gburrek,deegen,haeb\}@nt.upb.de}


\usepackage{comment}

\begin{document}

\robustify\bfseries
\sisetup{detect-weight=true,detect-inline-weight=math}

\maketitle

\begin{abstract}
We propose a spatio-spectral, combined model-based and data-driven diarization pipeline consisting of TDOA-based segmentation followed by embedding-based clustering. 
The proposed system requires neither access to multi-channel training data nor prior knowledge about the number or placement of microphones. It works for both a compact microphone array and distributed microphones, with minor adjustments.
Due to its superior handling of overlapping speech during segmentation, the proposed pipeline significantly outperforms the single-channel pyannote approach, both in a scenario with a compact microphone array and in a setup with distributed microphones.
Additionally, we show that, unlike fully spatial diarization pipelines, the proposed system can correctly track speakers when they change positions.
\end{abstract}

\section{Introduction}
Diarization systems assign regions of speech activity to the individual participants of a conversation, thus answering the question \enquote{Who spoke when?}. 
Essentially, they solve two tasks, segmentation and speaker assignment. The first is on identifying regions (segments) of constant speaker activity, while the second assigns speaker labels to each segment. There exists a large variety of methods for how these tasks are solved \cite{08_araki_spatial_dia, 18_snyder_xvector, 
22_horiguchi_eda_eend,
20_medennikov_tsvad, 23_plaquet_pyannote}. 

Here, we categorize diarization systems according to whether they use spectral or spatial cues, or both. 
Early diarization systems using spectral information employed statistical models \cite{06_Tranter},
% such as full-covariance Gaussians \cite{06_Tranter}, 
while recent systems rely on speaker embeddings, e.g., x-vectors or d-vectors \cite{18_snyder_xvector,17_Li_dvector,20_Desplanques_ecapa_tdnn}, extracted from audio segments, which are then clustered. Alternatively, they are used to directly predict the speech activity of all participants in a conversation on a frame-by-frame basis as in \gls{EEND} systems \cite{22_horiguchi_eda_eend}. 

If multi-channel input is available, spatial cues have been shown to deliver strong diarization results \cite{08_araki_spatial_dia, 07_anguera_spatial_dia_bf, 12_Ishiguro_prob_spat_dia, 16_Fakhry_prob_dia_spat}. In particular, they excel over spectral systems in regions of overlapping speech  \cite{wang2022spatial, 23_gburrek_asilomar}. However, one should be aware of the fact that segments of speech activity are assigned to positions or directions in space, rather than to speakers, with the consequence that speaker movements or speaker position changes can confuse the system.
Additionally, strong reflections can result in so-called phantom positions, indicating activity from a direction, where actually no speaker is present. It is also known that the quality of spatial cues depends on the inter-microphone distance, with reduced informativeness if this distance is small \cite{23_schmalen_libriwasn}.

There are only few examples of systems that use both spectral and spatial cues. 
Multi-channel information is used as auxiliary input of an otherwise spectral diarization system  in \cite{24_taherian_ssnd, 24_ustc_chime8_dia, 22_Zheng_tdoa_aug_dia}, leading to improved performance. However, this approach requires a dedicated training phase with multi-channel data. That this is a significant impediment became clear in the recently concluded NOTSOFAR-1 challenge, where the lack of in-domain training data was cited as the main reason why only few systems made explicit use of spatial information for diarization \cite{25_abramovski_notsofar_summary}. An example of a deeper integration of spectral and spatial information is the integrated model of \cite{24_cordlandwehr_integration}.

In this work, we introduce an alternative spatio-spectral diarization system. It shares similarities with the well-known pyannote diarization system, which is a purely spectral system that consists of (temporally) local segmentation followed by embedding-based global clustering \cite{23_plaquet_pyannote}. We propose to do segmentation with spatial features instead, using a model-based approach. With the local segmentation being strictly decoupled from the single-channel, embedding-based clustering stage, the proposed system does not require in-domain training data.

The spatial segmentation model is based on \cite{23_gburrek_asilomar}. It employs \glspl{TDOA} estimates to detect segments of speech activity for all active sources. 
Then, beamforming is applied to all segments with speech activity to enhance the target speaker, and suppress crosstalk in regions of overlapping speech.
Next, a  speaker embedding extractor is applied to the enhanced speech segments, and global clustering of embedding vectors is carried out to obtain the speaker assignments for all speech segments in the meeting.  This spectral clustering 
stage diminishes the impact of phantom positions, because embedding vectors computed from a segment representing a strong reflection will exhibit strong similarity with the segment containing the direct path signal of that speaker, such that they will be merged during clustering.
In this way, the advantages of both spectral and spatial processing are exploited, while mitigating their drawbacks: The spatial processing addresses noise and overlapping speech, while the spectral processing can cope with possible position changes of a speaker.

Unlike  \cite{23_gburrek_asilomar}, which requires globally constant speaker positions, the proposed system requires a speech source to be not moving only for a single segment of speech.
%, since segments of activity were grouped by a global, TDOA-based clustering. 
In contrast to \cite{24_taherian_ssnd}, the system does not require multi-channel training data, and it is independent of the number of microphones and its geometric arrangement. In the experiments, we show that it delivers good results both for a compact microphone array and for distributed microphones, with minimal adjustment of parameters.

\Cref{sec:dia} describes the proposed spatio-spectral diarization pipeline, which is evaluated in \cref{sec:experiments} both in a distributed and compact microphone setup in terms of \gls{DER} and \gls{WER}. Finally, \cref{sec:summary} offers some conclusions and 
an outlook on future work.

\section{Spatio-spectral diarization pipeline}
\label{sec:dia}

\begin{figure*}[bt]
    \centering
    \input{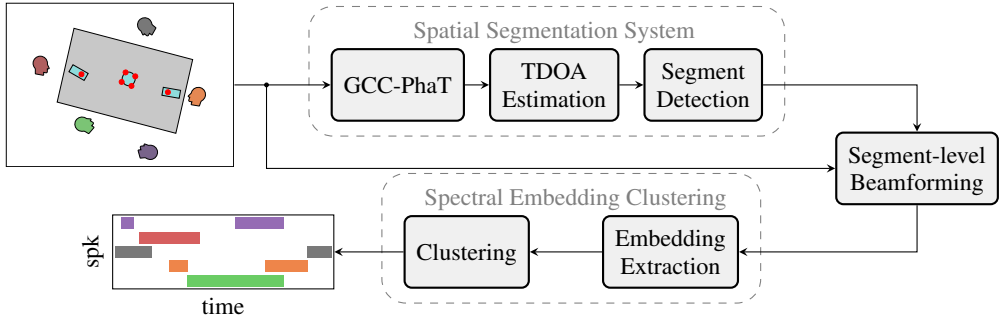}
    \vspace{-0.4em}
    \caption{Illustration of the proposed spatio-spectral diarization pipeline.}
    \label{fig:spatio_spectral_pipeline}
\end{figure*}
A multi-channel recording of a  meeting with $K$ speakers can be modeled in the time-frequency domain as the summation 
\begin{align}
    y_c(t,f) = \sum_{l(k)} s_{l(k)}(t,f) h_{c,l(k)}(t,f)
\end{align}
of delayed, clean speech signals $s_{l(k)}(t,f)$ which are padded to match the length of the conversation. Here,
$h_{c,l(k)}(t,f)$ denotes the acoustic transfer function between speaker $k$ and microphone $c$ for each speech segment $l(k)$ in time frame $t$ and frequency bin $f$. 

The proposed system cascades a TDOA-based local segment detection stage with a global spectral embedding-based clustering stage.
\cref{fig:spatio_spectral_pipeline} illustrates this pipeline.

\subsection{Multi-source TDOA estimation}
The \gls{TDOA} $\tau_{i,j}$ of a signal between two microphones $i$ and $j$ for a single active sound source can be found as the position of the maximum of the \gls{GCC-PhaT} \cite{76_Knapp_GCC_PhaT}
\begin{align}
    \tau_{i,j} = \arg\max\left( \mathrm{IFFT}\left(\frac{y_i(t,f)y_j(t,f)^*}{| y_i(t,f)y_j(t,f)^*|}\right)\right).
\end{align}
All pairwise \gls{TDOA} estimates are gathered in a \gls{TDOA} vector 
\begin{align}
\boldsymbol{\tau}= [\tau_{0,1}, \tau_{0,2},\tau_{1,2},\dots,\tau_{C-1,C}]^\mathrm{T},
\end{align}
 where $C$ is the total number of microphones.
\Gls{GCC-PhaT} exhibits one local maximum per source. Since multiple sources can be active at the same time,  $P$ local maxima are chosen as possible delays 
$\tau_p$, $1\le p\le P$
\cite{23_gburrek_asilomar}.
Thus, $P^{\frac{(C-1)C}{2}}$ different \gls{TDOA} vectors can be constructed from the estimated delays by combining all individual \glspl{TDOA}, while only up to $K$ vectors are physically grounded. 
To address this, a \gls{TDOA} vector is kept only 
if the sum of the delays over each closed loop of microphones is close to zero, e.g.,
\begin{align*}
    \tau_{0,1} +\tau_{1,2}+\tau_{2,0} < \tau_{\mathrm{th}},
\end{align*}
where the threshold $\tau_{\mathrm{th}}$ is set to a small value to account for numerical errors \cite{08_Scheuing_cdc, 23_gburrek_asilomar}. 
By performing this \gls{TDOA} estimation for each time frame, a set of \gls{TDOA} vectors and corresponding frame indices of speech activity is estimated, which are next grouped into speech segments.

\subsection{Temporally constrained segment detection}
The speech segment detection is performed as in \cite{23_gburrek_asilomar} by a temporally constrained leader-follower clustering. Here, the individual segments are determined by the pairwise Euclidean distance between all \gls{TDOA} vectors.
Two \gls{TDOA} vectors $\boldsymbol{\tau}_i$ and $\boldsymbol{\tau}_j$ can only belong to the same cluster if they do not exceed a maximal Euclidean distance $\Delta \tau_{\mathrm{max}}$, and if the temporal distance between frames $t_i$ and $t_j$ is smaller than \SI{1}{\second}.
This allows individual segments to contain short regions of either silence or where no \gls{TDOA} could be detected and prevents the formation of too large segments. Additionally, it prevents two consecutive segments from being merged since small segments are favored. 
After clustering, the detected segments $\hat{l}$ are specified by their start and end frames $t_{\mathrm{on},\hat l}$ and $t_{\mathrm{off},\hat l}$ as well as their median \gls{TDOA} vector $\bar{\boldsymbol{\tau}}_{\hat{l}}$, while their respective speaker labels are yet unknown.

\subsection{Segment-level beamforming}
According to the W-disjoint orthogonality property of speech \cite{02_Rickard_wdo}, each time-frequency bin (tf-bin) can be modeled to be either populated by a single source or by noise. 
This assumption underlying mask-based beamforming \cite{10_souden_mvdr} is used to estimate binary masks to perform segment-wise beamforming.

First, for each processed segment the tf-bins containing only noise are estimated. This is done according to \cite{19_Yang_multisource_localization} via the eigenvalue gap of the tf-wise \gls{SCM} estimates of the observation vector $\mathbf{y}(t,f)=(y_1(t,f), \ldots, y_C(t,f))^T$, which are gathered by averaging the outer product of the observation vector over a small local context.
Since these tf-wise \gls{SCM} estimates depict a dominant eigenvalue only for speech regions, bins are assigned to the noise mask if the eigenvalue gap between the first and second largest eigenvalue is below a threshold.

All remaining tf-bins are assigned to the mask of the processed segment or the interfering segments as follows.
First, \enquote{prototype} \glspl{SCM} are computed as the outer product
of the steering vectors $\mathbf{a}_{\hat{l}}$ corresponding to each $\boldsymbol{\tau}_{\hat{l}}$ %\cite{tdoa_to_steering} 
\begin{align}
    \boldsymbol\Phi_{\hat{l}} = \mathbf{a}_{\hat{l}}\mathbf{a}_{\hat{l}}^{\mathrm{H}}.
\end{align}
These prototypes
are compared against the instantaneous matrix of pairwise phase terms of the observation 
\begin{align}
    \boldsymbol \Psi = \frac{\mathbf{y}(t,f)\mathbf{y}(t,f)^\mathrm{H}}{| \mathbf{y}(t,f)\mathbf{y}(t,f)^\mathrm{H} |}
\end{align}
using the spatial covariance distance measure from \cite{05_Herdin_scm_dist}.  The binary mask of each speech segment is now formed by those tf-bins, whose \gls{SCM} is closest to the same prototype \gls{SCM}.

The segment-forming process can also result in superfluous segments that are caused by phantom positions. 
Before beamforming, these segments need to be identified and removed to prevent using a speaker's own reflection as an interferer during beamforming. 
Reflections are characterized by the fact that its \gls{SCM} shows a stronger deviation from the prototype \gls{SCM} at high frequencies, which is caused by phase errors so that the corresponding binary masks are sparsely populated for higher frequencies.
Therefore, the average mask activity between \SIrange{150}{3500}{\hertz} is compared against a threshold to determine whether the activity matches that of a speech signal. 
Segments containing too little activity are declared as caused by reflections and discarded.

Finally, the binary masks of the remaining segments need to be refined to fill up missing tf-bins that were assigned to discarded segments.
This can either be done by repeating the mask estimation on the reduced set of segments, or by using a \gls{cACGMM} for mask refinement as proposed in \cite{23_gburrek_asilomar, 24_ustc_chime8_dia}.
In the latter approach, 
a statistical mixture model is fitted to the data, which is initialized with the binary masks.
This refinement step is of low computational complexity because only few iterations are needed and because the model is applied to a single segment and not the whole meeting.

\subsection{Embedding extraction \& clustering}
For each beamformed segment,  the ResNet34-based d-vector model from \cite{24_boeddecker_speaker_reassignment} is employed to extract a speaker embedding. 
Then, HDBSCAN \cite{15_campello_hdbscan}, a hierarchical, density-based clustering approach is applied to the  embeddings.
Here, similar speaker embeddings are grouped by the pairwise cosine distance between them. 
Additionally, HDBSCAN marks outliers. These outlier segments are  merged into the most similar cluster. 
If two intersecting segments are assigned to the same cluster, the activity of both segments is merged. 
This allows merging phantom positions caused by reflections that were not detected in the beamforming stage.
Since this step employs spectral information only, embeddings of a speaker changing their position can still be merged into the same cluster.

\section{Experiments}
\label{sec:experiments}
\subsection{Experimental Setup}
For evaluation, the proposed pipeline is applied to the LibriCSS \cite{20_chen_libricss} and LibriWASN \cite{23_schmalen_libriwasn} data sets.
LibriCSS consists of re-recordings of simulated LibriSpeech \num{8}-speaker meetings ranging from \SIrange{0}{40}{\percent} overlapping speech with a duration of \SI{10}{\minute}. 
LibriWASN is an additional re-recording of the same synthetic meetings as LibriCSS, albeit in a distributed setup with multiple recording devices in two different rooms, exhibiting a T60 time of \SI{200}{\milli\second} (LibriWASN$_{200}$) and \SI{800}{\milli\second} (LibriWASN$_{800}$). 
% respectively.

The diarization pipeline is applied to \num{4} microphone channels, which is the smallest possible number for \gls{TDOA}-based source localization.
%in the given scenario. 
In the compact setup, the 4-element microphone array \textit{asnupb7} is used for LibriWASN, and \num{4}  of the non-center microphones in LibriCSS. For a distributed setup, four smartphones of  LibriWASN, the two \textit{Pixel6}, one \textit{Pixel7} and a \textit{Xiaomi} device, are used, and all channels are assumed to be synchronized both in terms of \gls{SRO} and \gls{STO}. 

The delay thresholds $\tau_{\mathrm{th}}$ are set to \num{1} and \num{2}
for the compact and distributed microphone setup, respectively, and
the maximum delays during segment detection $\Delta \tau_{\mathrm{max}}$ are set to \num{1} and \num{0.75} samples, to accommodate for the very different inter-microphone distances.
All remaining parameters are chosen independently of the scenarios, which encompass three different rooms and five different microphone setups.\footnote{\mbox{\url{github.com/fgnt/spatiospectral\_diarization}}} 

In addition to the \gls{DER} as a performance measure, the transcription performance of the downstream ASR system from \cite{watanabe2020PretrainedASR}  is evaluated in terms of \gls{cpWER}.
To this end, \gls{GSS} \cite{18_boeddeker_gss} is applied as in \cite{23_gburrek_asilomar} to extract the speech sources. 
For the ASR experiment,  all \num{7} microphone channels of LibriCSS are used to be comparable with the literature, while only  \num{4} channels are used for diarization. For calculating the \gls{DER} according to \cite{dscore}, no forgiveness collar is used, and the \gls{cpWER} is obtained using the \textit{meeteval} toolkit \cite{23_von_neumann_meeteval}.

\begin{table*}[bt]
    \centering
    \caption{DER and WER Performance of the proposed pipeline (with cACGMM refinement) for a distributed and compact setup. 
    }
    \label{tab:diarization_performance}
    \begin{tabular}{l l c c c c c c | c c c}
    \toprule
        Setup & Database & 0S & 0L & OV10 & OV20 & OV30 & OV40 & DER$_{\text{avg}}$ & DER$_{\mathrm{OV}}$ & cpWER\\
         \midrule 
                \multirow{2}{*}{Distributed}  & LibriWASN$_{200}$ &  3.46 & 3.90 & 3.30 & 3.92& 3.98 & 4.10 & 3.79 & 4.19 & 3.36\\% /scratch/hpc-prf-nt1/cord/models/libriwasn_spatiospectral/misleading_lavender_horse
          & LibriWASN$_{800}$ & 2.70 & 3.71 & 3.11 & 3.90 & 5.17 & 4.53 & 3.92 & 5.28 & 3.60 \\  %/scratch/hpc-prf-nt1/cord/models/libriwasn_spatiospectral/forthcoming_amber_flamingo/gss_results_libriwasn800
         \midrule 
         \midrule
                     & LibriWASN$_{200}$ & 3.52 & 3.81  & 5.34 & 4.93 & 5.57 & 6.89 & 5.16 & 7.00 & 5.13\\ %/scratch/hpc-prf-nt1/cord/models/libriwasn_spatiospectral/terrible_emerald_sturgeon
         Compact     & LibriWASN$_{800}$& 3.08 & 4.59 & 4.24 & 4.99 & 6.38 & 6.09 & 5.00 & 7.08 & 5.50\\ % /net/vol/cord/models/libriwasn_spatiospectral/intensive_azure_pike/gss_results_libriwasn800/sep_sigs/per_utt_fixed.json
                     & LibriCSS & 5.87 & 5.90 & 5.90 & 7.46 & 8.16 & 8.79 & 7.17 & 9.97 & 6.53  \\ % /scratch/hpc-prf-nt1/cord/models/libriwasn_spatiospectral/persistent_harlequin_peafowl/gss_results_libricss
         \bottomrule
    \end{tabular}
\end{table*}

\subsection{Diarization performance}
%\subsection{Diarization performance in distributed and compact microphone setups}
First, the proposed pipeline is evaluated w.r.t. its capability to be employed both in a distributed and a compact microphone setup.
\Cref{tab:diarization_performance} shows that, in the distributed setup of LibriWASN$_{200}$, the proposed system can perform diarization equally well in overlap and in single-speaker regions, achieving an average \gls{DER} of \SI{3.78}{\percent} and of \SI{4.19}{\percent} when only evaluating regions of overlapping speech. 
For the LibriWASN$_{800}$ database, the system still can achieve similar average ($\mathrm{DER}_{\mathrm{avg}}$) and overlap-\glspl{DER} (DER$_{\mathrm{OV}}$) of \SI{3.92}{\percent} and \SI{5.28}{\percent}, respectively. Compared to other systems like \cite{21_raj_scov} and \cite{24_cordlandwehr_geodesic} that try to address overlapping speech on a fully spectral level and achieve a DER$_{\mathrm{OV}}$ of \SIrange{25}{30}{\percent} for single-channel processing, this underlines the advantage of dedicated multi-channel processing to handle overlapping speech in diarization.

When switching to a compact scenario, the total performance decreases by \SIrange{1}{2}{\percent} absolute in terms of \gls{DER} and \gls{WER}, which is to be expected since the spatial cues used for \gls{TDOA} estimation become less informative and speakers are harder to separate. 
Still, the system is able to consistently obtain similar \glspl{DER} in single-speaker and overlap regions.

\subsection{Comparison to other systems}
\Cref{tab:der_comparison} compares the spatio-spectral pipeline against other systems in the compact microphone setup, which is the more common application for multi-channel meeting processing. 
We compared with the embedding-based, overlap-aware diarization system from \cite{21_raj_scov,23_raj_gpugss}  and the state-of-the-art,  hybrid diarization and enhancement system SSND \cite{24_taherian_ssnd} on LibriCSS.

To have a comparison also for LibriWASN, we implemented the following spatial and spectral systems as references:  a spatial-only pipeline directly clustering the median \gls{TDOA} vectors of the detected segments using single-linkage agglomerative clustering with outlier rejection,
and the single-channel pyannote 3.1 pipeline without any further modifications. 

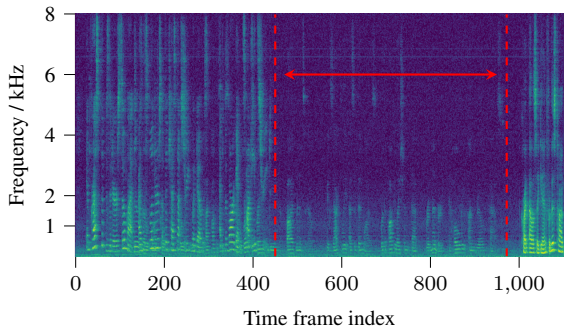
\begin{figure}[bt]
    \centering
    \setlength{\figureheight}{4.8cm}
    \setlength{\figurewidth}{8cm}
    % This file was created with tikzplotlib v0.10.1.
\begin{tikzpicture}

\definecolor{darkgray176}{RGB}{176,176,176}

\begin{axis}[
colormap/viridis,
font=\footnotesize,
height=\figureheight,
point meta max=8.93184019051793,
point meta min=-51.0681598094821,
tick align=outside,
tick pos=left,
width=\figurewidth,
x grid style={darkgray176},
xlabel={Time frame index},
xmin=-0.5, xmax=1099.5,
xtick style={color=black},
ytick = {64, 128, 256,384, 512},
yticklabels={1,2,4,6,8},
y grid style={darkgray176},
ylabel={Frequency / \si{\kilo\hertz}},
ymin=-0.5, ymax=512.5,
ytick style={color=black}
]
\addplot graphics [includegraphics cmd=\pgfimage,xmin=-0.5, xmax=1099.5, ymin=-0.5, ymax=512.5] {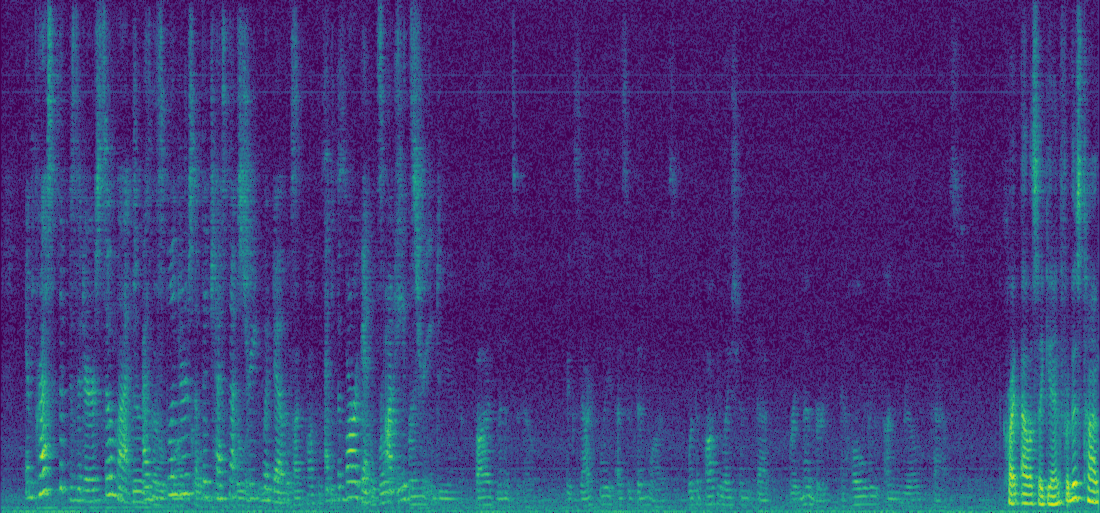};
\addplot[color=red, thick, dashed] table[row sep = crcr]{450 0 \\ 450 511 \\};
\addplot[color=red, thick, dashed] table[row sep = crcr]{972 0 \\ 972 511 \\};
\draw[red,<->, thick] (axis cs: 950,384) -- (axis cs: 470, 384);

\end{axis}

\end{tikzpicture}
    \vspace{-0.6em}
    \caption{Spectrogram of a LibriCSS segment with a single speech region of the defective loudspeaker (highlighted in red).}
    \label{fig:low_pass_spectrogram}
\end{figure}

It can be seen that the proposed spatio-spectral system outperforms both a purely spectral and spatial approach. This shows that the proposed system effectively combines both systems' advantages. Here, the spectral system shows stable, but lower performance due to solely using single-channel information, while the spatial model shows higher errors due to a coarser resolution and reflections in the environment.

On LibriWASN, the proposed system even proves slightly better when omitting the \gls{cACGMM}, demonstrating good performance even without additional segment refinement before beamforming.
However, for LibriCSS, which is comparable in difficulty and acoustic properties to LibriWASN$_{200}$, unexpectedly high error rates occur without the \gls{cACGMM} refinement.

After a closer analysis, these errors could be traced back to a single loudspeaker used during the recordings exhibiting a low-pass characteristic. This loudspeaker significantly attenuates frequencies above \SI{1.5}{\kilo\hertz}, as can be seen in \cref{fig:low_pass_spectrogram}.
Therefore, the filtering stage aimed at identifying reflections through the energy distribution of a speech signal inadvertently removes this loudspeaker's signal before embedding extraction. 
The \gls{cACGMM}-based mask refinement mitigates this effect and fills in gaps in the estimated speech activity, but cannot compensate for completely missed segments.
With refinement, the proposed system approaches the results of the state-of-the-art SSND system \cite{24_taherian_ssnd} on LibriCSS. While SSND also employs a spatio-spectral approach, it employs a fully data-driven model to cascade diarization and separation. Due to the transformer-based multi-channel \gls{EEND} network used for diarization, it requires matching training data on the corresponding microphone array geometry.  
Compared to it, the proposed model only requires training on VoxCeleb \cite{19_Nagrani_VoxCeleb} for the embedding extractor and is agnostic to microphone placement and geometry.

\begin{table}[bt]
    \sisetup{round-precision=1,round-mode=places, table-format = 2.1}
    \centering
    \setlength{\tabcolsep}{2pt}
    \caption{Comparison of the proposed pipeline to other systems in a compact microphone setup.\vspace{-0.4em}}
    \label{tab:der_comparison}
    \begin{tabular}{l S S S S S S}
    \toprule
        \multirow{2}{*}{System} & \multicolumn{2}{c}{{LibriWASN$_{200}$}} & \multicolumn{2}{c}{{LibriWASN$_{800}$}} & \multicolumn{2}{c}{{LibriCSS}} \\
        \cmidrule(lr){2-3}\cmidrule(lr){4-5}\cmidrule(lr){6-7}

         & {DER} & {WER} & {DER} & {WER} & {DER} & {WER} \\
         \midrule 
          SC + OV \cite{23_raj_gpugss} & {--} & {--} & {--}& {--} & 11.34 & 12.12\\
          % TS-VAD \cite{} & -- & -- & -- & -- & \\
          SSND \cite{24_taherian_ssnd} & {--} & {--} & {--} & {--} & 4.68 & 5.13 \\
          pyannote\footnotemark[2] & 12.75 & 8.84 & 12.56& 11.62 & 14.07& 10.22\\
          Spatial  & 11.52 & 4.63 & 12.29 & 6.47 & 15.80  & 9.81 \\ % 
          \midrule
          Proposed &  4.49 & 4.08 & 5.19 & 5.11  & 9.89 & 8.88  \\ % 
           ~~ + cACGMM & 5.16 & 5.13 & 5.00 & 5.50 & 7.17 & 6.52 \\
         \bottomrule
    \end{tabular}
    \label{tab:compact_der}
\end{table}
\footnotetext[2]{Applied to the first microphone channel using pyannote 3.1 \cite{23_plaquet_pyannote}}

\subsection{Spatio-spectral diarization for changing positions}
So far, all evaluations consider a static scenario with constant speaker positions.  
To verify the proposed, spatio-spectral pipeline's robustness against position changes, a \enquote{semi-static} LibriCSS and LibriWASN scenario
%, which we call "Musical Chairs",  
is created as follows. 
Two 
%successive 
meetings are concatenated so that each overlap subset (0S -- OV40) consists of \SI{20}{\minute} long meetings, where the speaker positions change after the first half of the meeting. 
Here, on average five out of the eight speakers are replaced by new speakers, while three speakers change location, resulting in meetings with \num{12}-\num{15} different speakers but only \num{8} speaker positions. This allows checking the system's performance both in case of different sources from the same location and speakers who change their position during the meeting.

\Cref{tab:music_libricss} shows that the proposed system, similar to the fully spectral pyannote pipeline, only marginally degrades when switching from a per-meeting evaluation (\textit{pm}) to the semi-static scenario (\textit{chg}).
Compared to this, the spatial-only system, as expected, cannot handle this scenario, since neither speakers changing their position nor multiple speakers from the same position can be accurately detected solely through their \glspl{TDOA}.  

\begin{table}[bt]
    \sisetup{round-precision=1,round-mode=places, table-format = 2.1}
    \newcommand{\STAB}[1]{\begin{tabular}{@{}c@{}}#1\end{tabular}}
        \centering
    \caption{Performance of the proposed system in the \enquote{semi-static} setup. \enquote{pm} denotes the cpWER on the individual meetings, \enquote{chg} on the concatenated meetings.\vspace{-0.4em}}
    \label{tab:music_libricss}
    \setlength{\tabcolsep}{4pt}
    \begin{tabular}{@{}cl@{}S S S S S S}
    \toprule
        &System & \multicolumn{2}{c}{{LibriWASN$_{200}$}} & \multicolumn{2}{c}{{LibriWASN$_{800}$}} & \multicolumn{2}{c}{{LibriCSS}} \\
        \cmidrule(lr){3-4}\cmidrule(lr){5-6}\cmidrule(lr){7-8}
         && {pm}  &{chg} & {pm} & {chg} & {pm} & {chg} \\
         \midrule 
         \multirow{2}{*}{\rotatebox[origin=c]{90}{\textcolor{gray}{\footnotesize Dist.}}}&
         Spatial   & 3.05 & %/net/vol/cord/models/libriwasn_spatiospectral/misleading_lavender_horse
         24.73 & %/net/vol/cord/models/libriwasn_spatiospectral_mc/different_emerald_flamingo
         4.45 & %/net/vol/cord/models/libriwasn_spatiospectral/forthcoming_amber_flamingo
         27.61 & %/net/vol/cord/models/libriwasn_spatiospectral_mc/thorough_azure_aardvark
         {--}& {--} \\ 
         & Proposed  & 3.36 & %/net/vol/cord/models/libriwasn_spatiospectral/misleading_lavender_horse
         3.32 & % /net/vol/cord/models/libriwasn_spatiospectral_mc/different_emerald_flamingo
         3.60 & % /net/vol/cord/models/libriwasn_spatiospectral/forthcoming_amber_flamingo
         4.02 & % /net/vol/cord/models/libriwasn_spatiospectral_mc/thorough_azure_aardvark
         {--} & {--} \\
         \midrule
         \midrule
        \multirow{3}{*}{\rotatebox[origin=c]{90}{\textcolor{gray}{\footnotesize Compact}}} &
        Spatial & 4.63 & %/net/vol/cord/models/libriwasn_spatiospectral/terrible_emerald_sturgeon
        74.21 & %/net/vol/cord/models/libriwasn_spatiospectral_mc/confidential_emerald_warbler/
        6.47 & %/net/vol/cord/models/libriwasn_spatiospectral/intensive_azure_pike
        73.70 & %/net/vol/cord/models/libriwasn_spatiospectral_mc/forward_tomato_goose
        9.81 & %/net/vol/cord/models/libriwasn_spatiospectral/persistent_harlequin_peafowl/
        29.80 \\ % /net/vol/cord/models/libriwasn_spatiospectral_mc/emotional_copper_cobra
         & pyannote \footnotemark[2] & 8.84 & 10.21 & 11.62 & 11.06 & 10.22 & 12.39 \\ % /net/vol/cord/models/pyannote_speaker_diarization_libri_musical_chair/3 DERs: LibriWASN200 12.82, LibriWASN800 12.30 LibriCSS 14.00  
         & Proposed & 5.13 & %/net/vol/cord/models/libriwasn_spatiospectral/terrible_emerald_sturgeon
         5.25 & % /net/vol/cord/models/libriwasn_spatiospectral_mc/confidential_emerald_warbler
         5.50 &  %/net/vol/cord/models/libriwasn_spatiospectral/intensive_azure_pike/
         5.61 & % /net/vol/cord/models/libriwasn_spatiospectral_mc/forward_tomato_goose/
         6.53 & %/net/vol/cord/models/libriwasn_spatiospectral/persistent_harlequin_peafowl/
         6.48 \\ %/net/vol/cord/models/libriwasn_spatiospectral_mc/emotional_copper_cobra/
         \bottomrule
    \end{tabular}
\end{table}

\section{Conclusion}
\label{sec:summary}
In this work, we presented an approach that combines spatial segmentation with a spectral, embedding-based clustering model for the diarization of meetings without requiring in-domain training data.
The proposed model can be deployed in compact and distributed microphone setups without large performance differences and with only minimal parameter changes.
It was shown to robustly handle regions of overlapping speech and speaker position changes.
Because the spatial subsystem is model-based instead of data-driven, it does not require in-domain training data and prior knowledge about the microphone configuration.

In future work, we will focus on completely integrating the speech enhancement stage into the segment-level beamforming of the spatio-spectral pipeline and extend the \gls{TDOA} segmentation to handle continuous speaker movements.

\ifinterspeechfinal
    \section{Acknowledgements}
Computational Resources were provided by BMBF/NHR/PC2.
\fi

\bibliographystyle{IEEEtran}
\bibliography{mybib}

\begin{thebibliography}{10}
\providecommand{\url}[1]{#1}
\csname url@samestyle\endcsname
\providecommand{\newblock}{\relax}
\providecommand{\bibinfo}[2]{#2}
\providecommand{\BIBentrySTDinterwordspacing}{\spaceskip=0pt\relax}
\providecommand{\BIBentryALTinterwordstretchfactor}{4}
\providecommand{\BIBentryALTinterwordspacing}{\spaceskip=\fontdimen2\font plus
\BIBentryALTinterwordstretchfactor\fontdimen3\font minus
  \fontdimen4\font\relax}
\providecommand{\BIBforeignlanguage}[2]{{%
\expandafter\ifx\csname l@#1\endcsname\relax
\typeout{** WARNING: IEEEtran.bst: No hyphenation pattern has been}%
\typeout{** loaded for the language `#1'. Using the pattern for}%
\typeout{** the default language instead.}%
\else
\language=\csname l@#1\endcsname
\fi
#2}}
\providecommand{\BIBdecl}{\relax}
\BIBdecl

\bibitem{08_araki_spatial_dia}
S.~Araki, M.~Fujimoto, K.~Ishizuka, H.~Sawada, and S.~Makino, ``A {DOA} based
  speaker diarization system for real meetings,'' in \emph{Hands-Free Speech
  Communication and Microphone Arrays (HSCMA)}, 2008, pp. 29--32.

\bibitem{18_snyder_xvector}
D.~Snyder, D.~Garcia-Romero, G.~Sell, D.~Povey, and S.~Khudanpur, ``X-vectors:
  Robust {DNN} embeddings for speaker recognition,'' in \emph{Proc. IEEE
  ICASSP}, 2018, pp. 5329--5333.

\bibitem{22_horiguchi_eda_eend}
S.~Horiguchi, Y.~Fujita, S.~Watanabe, Y.~Xue, and P.~Garcia, ``Encoder-decoder
  based attractors for end-to-end neural diarization,'' \emph{IEEE/ACM
  Transactions on Audio, Speech, and Language Processing}, vol.~30, pp.
  1493--1507, 2022.

\bibitem{20_medennikov_tsvad}
I.~Medennikov, M.~Korenevsky, T.~Prisyach, Y.~Khokhlov, M.~Korenevskaya
  \emph{et~al.}, ``Target-speaker voice activity detection: A novel approach
  for multi-speaker diarization in a dinner party scenario,'' in \emph{Proc.
  ISCA Interspeech}, 2020, pp. 274--278.

\bibitem{23_plaquet_pyannote}
A.~Plaquet and H.~Bredin, ``{Powerset multi-class cross entropy loss for neural
  speaker diarization},'' in \emph{Proc. ISCA Interspeech}, 2023.

\bibitem{06_Tranter}
S.~Tranter and D.~Reynolds, ``An overview of automatic speaker diarization
  systems,'' \emph{IEEE Transactions on Audio, Speech, and Language
  Processing}, vol.~14, no.~5, pp. 1557--1565, 2006.

\bibitem{17_Li_dvector}
C.~Li, X.~Ma, B.~Jiang, X.~Li, X.~Zhang \emph{et~al.}, ``Deep speaker: An
  end-to-end neural speaker embedding system,'' \emph{arXiv preprint
  arXiv:1705.02304}, 2017.

\bibitem{20_Desplanques_ecapa_tdnn}
B.~Desplanques, J.~Thienpondt, and K.~Demuynck, ``{ECAPA-TDNN: Emphasized
  Channel Attention, Propagation and Aggregation in TDNN Based Speaker
  Verification},'' in \emph{Proc. ISCA Interspeech}, 2020, pp. 3830--3834.

\bibitem{07_anguera_spatial_dia_bf}
X.~Anguera, C.~Wooters, and J.~Hernando, ``Acoustic beamforming for speaker
  diarization of meetings,'' \emph{IEEE Transactions on Audio, Speech, and
  Language Processing}, vol.~15, no.~7, pp. 2011--2022, 2007.

\bibitem{12_Ishiguro_prob_spat_dia}
K.~Ishiguro, T.~Yamada, S.~Araki, T.~Nakatani, and H.~Sawada, ``Probabilistic
  speaker diarization with bag-of-words representations of speaker angle
  information,'' \emph{IEEE Transactions on Audio, Speech, and Language
  Processing}, vol.~20, no.~2, pp. 447--460, 2011.

\bibitem{16_Fakhry_prob_dia_spat}
M.~Fakhry, N.~Ito, S.~Araki, and T.~Nakatani, ``Modeling audio directional
  statistics using a probabilistic spatial dictionary for speaker diarization
  in real meetings,'' in \emph{IEEE International Workshop on Acoustic Signal
  Enhancement (IWAENC)}, 2016, pp. 1--5.

\bibitem{wang2022spatial}
J.~Wang, Y.~Liu, B.~Wang, Y.~Zhi, S.~Li \emph{et~al.}, ``Spatial-aware speaker
  diarization for multi-channel multi-party meeting,'' in \emph{Proc. ISCA
  Interspeech}, 2022.

\bibitem{23_gburrek_asilomar}
T.~Gburrek, J.~Schmalenstroeer, and R.~Haeb-Umbach, ``Spatial diarization for
  meeting transcription with ad-hoc acoustic sensor networks,'' in \emph{57th
  Asilomar Conference on Signals, Systems, and Computers}, 2023, pp.
  1399--1403.

\bibitem{23_schmalen_libriwasn}
J.~Schmalenstroeer, T.~Gburrek, and R.~Haeb-Umbach, ``{LibriWASN}: A data set
  for meeting separation, diarization, and recognition with asynchronous
  recording devices,'' in \emph{ITG conference on Speech Communication}, Sep
  2023.

\bibitem{24_taherian_ssnd}
H.~Taherian and D.~Wang, ``Multi-channel conversational speaker separation via
  neural diarization,'' \emph{IEEE/ACM Transactions on Audio, Speech, and
  Language Processing}, 2024.

\bibitem{24_ustc_chime8_dia}
R.~Wang, S.~Niu, G.~Yang, J.~Du, S.~Qian \emph{et~al.}, ``Incorporating spatial
  cues in modular speaker diarization for multi-channel multi-party meetings,''
  \emph{arXiv preprint arXiv:2409.16803}, 2024.

\bibitem{22_Zheng_tdoa_aug_dia}
N.~Zheng, N.~Li, J.~Yu, C.~Weng, D.~Su \emph{et~al.}, ``Multi-channel speaker
  diarization using spatial features for meetings,'' in \emph{Proc. IEEE
  ICASSP}, 2022, pp. 7337--7341.

\bibitem{25_abramovski_notsofar_summary}
I.~Abramovski, A.~Vinnikov, S.~Shaer, N.~Kanda, X.~Wang \emph{et~al.},
  ``Summary of the {NOTSOFAR-1} challenge: Highlights and learnings,''
  \emph{arXiv preprint arXiv:2501.17304}, 2025.

\bibitem{24_cordlandwehr_integration}
T.~Cord-Landwehr, C.~Boeddeker, and R.~Haeb-Umbach, ``Simultaneous diarization
  and separation of meetings through the integration of statistical mixture
  models,'' \emph{arXiv preprint arXiv:2410.21455}, 2024.

\bibitem{76_Knapp_GCC_PhaT}
C.~Knapp and G.~Carter, ``The generalized correlation method for estimation of
  time delay,'' \emph{IEEE Transactions on Acoustics, Speech, and Signal
  Processing}, vol.~24, no.~4, pp. 320--327, 1976.

\bibitem{08_Scheuing_cdc}
J.~Scheuing and B.~Yang, ``Disambiguation of {TDOA} estimation for multiple
  sources in reverberant environments,'' \emph{IEEE Transactions on Audio,
  Speech, and Language Processing}, vol.~16, no.~8, pp. 1479--1489, 2008.

\bibitem{02_Rickard_wdo}
S.~Rickard and O.~Yilmaz, ``On the approximate {W}-disjoint orthogonality of
  speech,'' in \emph{Proc. IEEE ICASSP}, vol.~1, 2002, pp. I--529--I--532.

\bibitem{10_souden_mvdr}
M.~Souden, J.~Benesty, and S.~Affes, ``On optimal frequency-domain multichannel
  linear filtering for noise reduction,'' \emph{IEEE Transactions on Audio,
  Speech, and Language Processing}, vol.~18, no.~2, pp. 260--276, 2010.

\bibitem{19_Yang_multisource_localization}
B.~Yang, H.~Liu, C.~Pang, and X.~Li, ``Multiple sound source counting and
  localization based on tf-wise spatial spectrum clustering,'' \emph{IEEE/ACM
  Transactions on Audio, Speech, and Language Processing}, vol.~27, no.~8, pp.
  1241--1255, 2019.

\bibitem{05_Herdin_scm_dist}
M.~Herdin, N.~Czink, H.~Ozcelik, and E.~Bonek, ``Correlation matrix distance, a
  meaningful measure for evaluation of non-stationary mimo channels,'' in
  \emph{IEEE 61st Vehicular Technology Conference}, vol.~1, 2005, pp. 136--140
  Vol. 1.

\bibitem{24_boeddecker_speaker_reassignment}
C.~Boeddeker, T.~Cord-Landwehr, and R.~Haeb-Umbach, ``Once more diarization:
  Improving meeting transcription systems through segment-level speaker
  reassignment,'' in \emph{Proc. ISCA Interspeech}, 2024, pp. 1615--1619.

\bibitem{15_campello_hdbscan}
R.~J. G.~B. Campello, D.~Moulavi, A.~Zimek, and J.~Sander, ``Hierarchical
  density estimates for data clustering, visualization, and outlier
  detection,'' \emph{ACM Trans. Knowl. Discov. Data}, 2015.

\bibitem{20_chen_libricss}
Z.~Chen, T.~Yoshioka, L.~Lu, T.~Zhou, Z.~Meng \emph{et~al.}, ``Continuous
  speech separation: Dataset and analysis,'' in \emph{Proc. IEEE ICASSP}, 2020,
  pp. 7284--7288.

\bibitem{watanabe2020PretrainedASR}
\BIBentryALTinterwordspacing
S.~Watanabe, ``{ESPnet2 pretrained model, Shinji Watanabe/librispe
  ech\_asr\_train\_asr\_transformer\_e18\_raw\_bpe\_sp\_valid .acc.best,
  fs=16k, lang=en},'' Jul. 2020. [Online]. Available:
  \url{https://doi.org/10.5281/zenodo.3966501}
\BIBentrySTDinterwordspacing

\bibitem{18_boeddeker_gss}
C.~Boeddecker, J.~Heitkaemper, J.~Schmalenstroeer, L.~Drude, J.~Heymann
  \emph{et~al.}, ``Front-end processing for the {CHiME}-5 dinner party
  scenario,'' in \emph{Proc. 5th International Workshop on Speech Processing in
  Everyday Environments (CHiME)}, 2018, pp. 35--40.

\bibitem{dscore}
O.~Sadjadi, C.~Greenberg, E.~Singer, L.~Mason, and D.~Reynolds,
  ``\BIBforeignlanguage{en}{{NIST} 2021 speaker recognition evaluation plan},''
  2021-07-12 04:07:00 2021.

\bibitem{23_von_neumann_meeteval}
T.~von Neumann, C.~Boeddeker, M.~Delcroix, and R.~Haeb-Umbach, ``{MeetEval}: A
  toolkit for computation of word error rates for meeting transcription
  systems,'' in \emph{Proc. 7th International Workshop on Speech Processing in
  Everyday Environments (CHiME)}, 2023, pp. 27--32.

\bibitem{21_raj_scov}
D.~Raj, Z.~Huang, and S.~Khudanpur, ``Multi-class spectral clustering with
  overlaps for speaker diarization,'' in \emph{IEEE Spoken Language Technology
  Workshop (SLT)}, 2021, pp. 582--589.

\bibitem{24_cordlandwehr_geodesic}
T.~Cord-Landwehr, C.~Boeddeker, C.~Zoril\u{a}, R.~Doddipatla, and
  R.~Haeb-Umbach, ``Geodesic interpolation of frame-wise speaker embeddings for
  the diarization of meeting scenarios,'' in \emph{Proc. IEEE ICASSP}, 2024,
  pp. 11\,886--11\,890.

\bibitem{23_raj_gpugss}
D.~Raj, D.~Povey, and S.~Khudanpur, ``{GPU}-accelerated guided source
  separation for meeting transcription,'' in \emph{Proc. ISCA Interspeech},
  2023, pp. 3507--3511.

\bibitem{19_Nagrani_VoxCeleb}
A.~Nagrani, J.~S. Chung, W.~Xie, and A.~Zisserman, ``{Voxceleb}: Large-scale
  speaker verification in the wild,'' \emph{Computer Science and Language},
  2019.

\end{thebibliography}

\end{document}